# Student understanding of Fermi energy, the Fermi-Dirac distribution and total electronic energy of a free electron gas

Paul Justice, Emily Marshman, and Chandralekha Singh

*Department of Physics and Astronomy, University of Pittsburgh, Pittsburgh, PA, 15260*

**Abstract.** We investigated the difficulties that physics students in upper-level undergraduate quantum mechanics and graduate students after quantum and statistical mechanics core courses have with the Fermi energy, the Fermi-Dirac distribution and total electronic energy of a free electron gas after they had learned relevant concepts in their respective courses. These difficulties were probed by administering written conceptual and quantitative questions to undergraduate students and asking some undergraduate and graduate students to answer those questions while thinking aloud in one-on-one individual interviews. We find that advanced students had many common difficulties with these concepts after traditional lecture-based instruction. Engaging with a sequence of clicker questions improved student performance, but there remains room for improvement in their understanding of these challenging concepts.

## INTRODUCTION

In the past two decades, many investigations have focused on improving student learning of quantum mechanics (QM); e.g., see Ref. [1-14] that focus on students' ideas about general QM issues. Some of these investigations focused on QM at the introductory level [3-4]. Others focused on energy related issues in QM, e.g., visualizing potential energy diagrams [1], a model of conductivity [5], quantum scattering and tunneling [6], and energy measurement [13]. Yet others have focused on student conceptions of broader QM issues [2,8,9,11], developing a curriculum involving interactive simulations [7,10], successful learner's epistemological framing of QM [12] and whether the way we teach QM is a lost opportunity for engaging motivated students [14]. Our group has been deeply involved in investigations to improve student learning of QM and using the research on student difficulties as a guide to develop research-validated learning tools [15-33]. For example, our earlier multi-university investigations [9,15,16] show that advanced undergraduates and graduate students have many common difficulties with QM regardless of institution, instructor or textbook and students also struggle in categorizing QM problems based upon similarity of solutions [17]. We used research as a guide to develop research-validated learning tools to help students learn about the Stern-Gerlach experiment [18], quantum measurement [19], addition of angular momentum [20], quantum key distribution [21], Larmor precession of spin [22], single photon QM including quantum eraser in the context of Mach-Zehnder interferometer [24,26], observables, probability distributions of outcomes of measurement and expectation values [27,28,29], the double slit experiment [30,31], and degenerate perturbation theory including Zeeman effect and fine structure of the hydrogen atom and identical particles [32,33]. We also reviewed student difficulties in QM, developed a framework for understanding student difficulties in QM, investigated the impact of learning from mistakes in QM [23] and investigated the diversity of physics instructors' attitudes and approaches to teaching QM [25].

**Goals of this investigation:** While the Fermi energy, the Fermi-Dirac distribution and total electronic energy of a free electron gas are important concepts taught in advanced quantum and statistical mechanics courses, there has been little work done on investigating student difficulties with these concepts [34]. In particular, the only earlier study on alternative conceptions of Fermi energy focused on teaching visualization of the Fermi-Dirac distribution function as a function of temperature using Microsoft Excel spreadsheets [34]. One goal of our investigation presented here was to probe the difficulties that upper-level physics undergraduates in a QM course and physics graduate students after quantum and statistical mechanics core courses have with these concepts after they had learned them in their respective courses. These difficulties were probed by administering written conceptual and quantitative questions to undergraduate students and asking some students in undergraduate and graduate courses to answer the questions while thinking aloud [35] in one-on-one interviews. Student difficulties with these concepts were investigated after traditional lecture-based instruction by administering a "pretest" which involved both conceptual and quantitative questions requiring understanding of the procedures on these topics. We also investigated the impact of a clicker question sequence (CQS) on these topics on undergraduate student performance by administering the same test that was administered as a "pretest" (after traditional instruction) also as a "posttest" because a related goal of the investigation was to evaluate how the ordering of different components - traditional instruction, pretest, CQS, posttest - impacted student performance on the pretest and posttest. Moreover, the CQS questions combined conceptual and quantitative knowledge because the CQS was developed and validated to help students both learn the relevant concepts as well as gain the ability to solve problems. The extent to which the students will benefit from a CQS that integrates conceptual and quantitative knowledge is unclear a-priori especially because these are advanced students. Therefore, one goal of the investigation was to evaluate the effectiveness of this type of a CQS that combines conceptual and quantitative knowledge of these advanced topics by measuring improvement from the pretest to posttest along these dimensions. The change in performance from the pretest to posttest can shed light on the expertise of advanced students and their ability to learn from this type of a CQS and transfer their learning to generate the posttest responses without any scaffolding support.

**Background on Relevant Topics:** The free electron gas [34] is a commonly taught model of solids (metals) in a two-semester upper-level undergraduate and core graduate quantum mechanics sequence. This model ignores many realities of real metals like the electron charge and the underlying lattice. The main consideration is the Pauli exclusion principle, which requires electrons to occupy distinct single-particle states. In this non-interacting fermionic model in three spatial dimensions, electrons can be considered to move freely in a three dimensional infinite rectangular box (although solid state physicists use periodic boundary condition). Since the size of a solid is macroscopic and the number of electrons is very large (of the order of Avogadro's number), the actual shape of the solid is not important for determining the bulk properties of the solid, and the free electron gas model explains many qualitative properties of conductors reasonably well [34].

Although the concepts of Fermi energy and density of states are defined more broadly, these are two key concepts advanced students often learn for the first time in the context of the free electron gas model of a solid. The Fermi energy $E_f$ is the energy of the highest occupied state at absolute zero temperature $T = 0\ K$. The density of states of the system $D(\epsilon)$ is the number of states per interval of energy for a given energy $\epsilon$. The concepts of Fermi energy and density of states can be used to calculate the total electronic energy of a solid at $T = 0\ K$, i.e., $E_{tot} = \int_0^{E_F} D(\epsilon) \epsilon\, d\epsilon$.

In quantum statistical mechanics [34], the concept of the distribution function, $n(\epsilon)$, which is defined as the average number of particles in a given single-particle state with energy $\epsilon$ at a given temperature $T$, becomes important. At $T = 0\ K$, the Fermi-Dirac (FD) distribution function for a non-interacting fermionic system, e.g., electrons discussed here, is a step function such that all single-particle states below the Fermi energy are completely filled and all states above the Fermi energy are empty. However, as the temperature increases, the probability of occupying higher single particle energy states increases. Moreover, at very high temperature when the de Broglie wavelength is very small, the wavefunctions of different electrons do not overlap. At this temperature, the Fermi-Dirac distribution function reduces to the Maxwell-Boltzmann (MB) distribution function, which is an exponential function of energy $\epsilon$. The Fermi-Dirac distribution function is $n_{FD}(\epsilon) = \frac{1}{e^{\frac{\epsilon - \mu(T)}{k_B T}} + 1}$, where $\mu$ is the chemical potential, defined as the energy required to add an extra electron to the system. The chemical potential $\mu$ depends on temperature $T$ and is equal to the Fermi energy at $T = 0\ K$. The total electronic energy of the system at temperature $T$ is given by $E_{tot} = \int_0^\infty n(\epsilon) D(\epsilon) \epsilon\, d\epsilon$. For comparison, the Bose-Einstein (BE) distribution function for a bosonic system is given by $n_{BE}(\epsilon) = \frac{1}{e^{\frac{\epsilon - \mu(T)}{k_B T}} - 1}$. There are no constraints on the number of bosons in a given single-particle state.

**Clicker Question Sequence on Fermi energy, Total Electronic Energy and Fermi-Dirac Distribution Function:** While clicker questions for introductory [36-38] and upper-level QM [39-40] have been developed, there have been very few efforts [41] toward a systematic development and implementation of clicker question sequences (CQSs), e.g., those on a given concept in which the questions build on each other effectively to help students organize their knowledge. Here we discuss how student performance on questions probing their understanding of the Fermi energy and total electronic energy of a free electron gas, and of the Fermi-Dirac distribution function, in an upper-level undergraduate QM course was impacted by a CQS focusing on these concepts. This CQS was developed and validated by contemplating the learning objectives, and by refining and fine-tuning existing clicker questions or developing new questions. The learning objectives related to Fermi energy include helping students learn to calculate the Fermi energy in terms of the free electron number density and realize that the Fermi energy is not an extensive quantity. The learning objectives related to the total electronic energy of a free electron gas include helping students learn to calculate the total electronic energy and realize that this quantity is extensive and therefore scales with the size of the system. The learning objectives related to the distribution functions include preparing students to be able to write an expression for them, be able to distinguish the Fermi-Dirac distribution function from the Bose-Einstein and Maxwell-Boltzmann distribution functions, be able to explain when the Fermi-Dirac (and Bose-Einstein) distribution functions will approach the Maxwell-Boltzmann distribution function, and be able to graphically represent the Fermi-Dirac distribution function at $T = 0\ K$ and at $T > 0\ K$.

We note that since we want students to be able to develop both conceptual knowledge and understanding of the procedures (e.g., be able to derive expressions for the Fermi energy, density of states and total electronic energy of a free electron gas), it is possible that using tutorial worksheets would have been more effective than using a CQS. However, due to logistical reasons, particularly because more instructors are using clickers in their courses, we focused on developing and validating a CQS which integrates conceptual and quantitative knowledge. Moreover, while a majority of clicker questions in the past [36-40] has only focused on conceptual questions, our goal was to evaluate the extent to which the CQS which integrates conceptual and quantitative knowledge will improve advanced students' performance on posttest on these topics. The extent to which advanced students will benefit from the CQS and be able to generate responses to open-ended questions in the posttest is unclear a-priori. We hypothesized that gaining systematic facility in deriving expressions for the Fermi energy, density of states and total electronic energy of a free electron gas through the CQS may improve student ability to make correct inferences in various situations. We also note that we did not want to cause cognitive overload for students by providing many incorrect choices when students were thinking through these challenging topics via the CQS. Therefore, we limited the number of incorrect choices involving student difficulties in the CQS but strongly encouraged instructors to discuss them after the clicker questions in the general class discussions that ensued. In particular, we hypothesized that the combined conceptual and quantitative questions in the CQS were already challenging for students and including many incorrect responses based upon student difficulties in the questions would cause cognitive overload for students and make it difficult for them to process the combined conceptual and quantitative knowledge in each question. The validation was an iterative process. The three authors met to holistically examine the instructional materials from the past few years on these topics in an upper-level undergraduate QM course at a large university, which included existing clicker questions on these concepts. In particular, the questions in the CQS were developed or adapted from prior clicker questions and sequenced to balance difficulties, avoid change of both concept and context between consecutive questions as appropriate in order to avoid a cognitive overload, and include a mix of abstract and concrete questions to help students develop a good grasp of the concepts. After the initial development of the CQS, we iterated the CQS with three physics faculty members who provided valuable feedback on fine-tuning and refining both the CQS as a whole and individual questions that were developed and adapted from existing clicker questions to ensure that the questions were unambiguously worded and build on each other based upon the learning objectives. We also conducted think-aloud interviews [35] with advanced students who had learned these concepts via traditional lecture-based instruction to ensure that they interpreted the CQS questions as intended and the sequencing of questions provided appropriate guidance to help them learn relevant concepts.

## METHODOLOGY

The students who participated in this study were upper-level physics undergraduates in a second semester junior/senior-level QM course and graduate students who had taken graduate core quantum and statistical mechanics courses. Both the undergraduate and graduate courses typically have 10-20 students each year. The undergraduate students had also taken the first semester of QM in the preceding semester and a majority of them had also taken or were concurrently taking a one-semester undergraduate thermodynamics and statistical mechanics course. The student difficulties were investigated by administering open-ended questions in written form to undergraduate students in the QM course after traditional lecture-based instruction in relevant concepts (we will call this pretest) and also after students had engaged with the CQS on relevant concepts (we will call this posttest). We wanted to investigate how the performance on the pretest and posttest compared with each other in order to gauge the extent to which the difficulties were reduced after the CQS compared to after traditional lecture-based instruction. As noted earlier, the CQS questions were validated with the help of physics instructors who had taught QM and/or statistical mechanics courses several times (the questions were iterated with them to ensure that they were robust and interpreted unambiguously by physics experts) and students to ensure, e.g., that they interpreted the questions as intended.

In addition to written tests, we also conducted individual semi-structured interviews with a subset of students in the undergraduate course and with graduate students after they had completed core graduate QM and statistical mechanics courses in which relevant concepts were covered. Individual interviews were conducted with 9 students (13 hours total) using a think-aloud protocol [35] to better understand the rationale for student responses. During the interviews, similar to the in-class written administration in the undergraduate course, students were first given the open-ended questions after traditional instruction (pretest), then they worked through the CQS, and then they were given the open-ended questions again as a posttest. The testing materials were developed and validated to assess student understanding of these concepts based upon the learning objectives delineated earlier. During these semi-structured interviews, students were asked to verbalize their thought processes while they answered the questions. They read the questions and answered them to the best of their ability without being disturbed. We prompted them to think aloud [35] if they were quiet for a long time. After students had finished answering a particular question to the best of their ability, we asked them to further clarify and elaborate issues that they had not clearly addressed earlier.

The final version of the CQS questions pertaining to this series on the free electron gas are shown in the Appendix. At T = 0 K, the CQS has 11 questions (first 11 questions as shown in the Appendix) but the only relevant questions for the pre/posttests are CQ1-CQ7 and CQ10 (note that CQ11 pertaining to the two dimensional free-electron gas was not administered to students in this study but it is included here as an instructional resource). At T > 0 K, the CQS has six questions (last six questions in the Appendix). Since the instructor used the textbook by Griffiths [34], the notation and discussion (e.g., about an octant) is consistent with that treatment although in solid state physics, periodic boundary conditions are used for application to transport properties and to extend the discussion to the band model (which is necessary for understanding band gaps and properties of systems other than metals, which the free electron model fails to do). The first section (CQ1-CQ10), which focuses on the Fermi energy, density of states, and total electronic energy of a free electron gas at absolute zero temperature, was administered in one class period. The second section focusing on the FD and BE distribution functions and their limiting behavior was administered in another class period. Unless specified otherwise, students were instructed to assume that the symbol for momentum $\hbar k$ signifies the magnitude of the momentum vector.

We note that the CQS was implemented with peer interaction [36-42] in the upper-level undergraduate QM class after traditional lecture-based instruction in relevant concepts on the Fermi energy, density of states, and total electronic energy of a solid within the free electron gas model and after learning about the Fermi-Dirac distribution function in the same QM course. In particular, after posing each question, the instructor asked students to talk to a peer before selecting their responses via clickers [36-41]. When students engaged with the CQS in one-on-one interviews, there was no peer discussion. Prior to engaging with the CQS, students took the pretest after traditional lecture-based instruction. After engaging with the CQS, they took the posttest. The five questions on the pretest and posttest focusing on the topics of Fermi energy, density of states, total electronic energy, and distribution functions are the same and they are given in Figure 1. In the individual interviews, students answered all five pre/posttest questions together before and after engaging with both sections of the CQS. However, in the written pretest in the undergraduate QM course, students were given the first two questions of the test together in a pretest part I after students learned about the Fermi energy, density of states and total electronic energy of the free electron gas via lecture and questions 3-5 together in another pretest (part II of the pretest) on a different day after students learned about the distribution functions via lecture-based instruction. The posttest was also administered in two parts on two days after students engaged with both sections of the CQS focusing on these concepts and questions 1,4 and 5 were deliberately administered together on one day (part I of the posttest) and questions 2 and 3 were administered together on another day (part II of the posttest). Students had sufficient time to answer the questions on the pre/posttests. We note that since questions 1 and 2 on the test are related (question 1 asks about the Fermi energy and total electronic energy as a conceptual question whereas question 2 asks about them as a question focused on mathematical manipulation) and questions 3-5 are related (question 3 seeks mathematical expressions for the distribution functions and constraints on the number of particles in each single-particle state whereas question 5 asks about the Fermi-Dirac distribution function in graphical representation at two different temperatures and question 4 asks about a limiting case of the quantum distribution functions), the grouping of the questions on the two parts of the *pretest* may have made it easier for the students to answer them in that the related questions in each part of the pretest can prime students to answer the questions more easily than on the *posttest* in which the different types of questions on Fermi energy, total electronic energy and Fermi-Dirac distribution function were deliberately mixed in parts I and II. For example, on pretest part I, if students answered question 2 correctly and found that the Fermi energy does not scale with the size of the system but the total electronic energy does, they can potentially use those results to answer the conceptual question 1 correctly. Similarly, on pretest part II, a student who wrote the correct expression for the Fermi-Dirac distribution function when explicitly prompted in question 3 can potentially take advantage of it to come up with its correct graphical representation in question 5 or find the correct limiting case in question 4. However, as we will see in the results section, student performance on all questions was poor on the pretest after traditional lecture-based instruction, and they did not benefit on the pretest from having similar questions grouped together.

> (1) On your desk, you have two cubes A and B which have N and 2N copper atoms, respectively.
>   a. At temperature $T = 0$ K, which cube has the higher Fermi energy? Explain your reasoning.
>   b. At temperature $T = 0$ K, which cube has the higher total electronic energy? Explain your reasoning.
> (2) Consider a system of N non-interacting electrons in a **_two-dimensional_** infinite square well of area A. Let σ be the number of free electrons per unit area and m be the mass of an electron. Using this information, do the following:
>   a. Find an expression for the _Fermi energy_ for the system.
>   b. Derive an expression for the density of states.
>   c. Find an expression for the _total electronic energy_ for the system.
> (3) Give an expression for each of the following distribution functions and give the constraints on each of these distribution functions in terms of how many particles each single-particle state can accommodate.
>   a. Maxwell-Boltzmann: $n_{MB}(\epsilon) =$
>   b. Fermi-Dirac: $n_{FD}(\epsilon) =$
>   c. Bose-Einstein: $n_{BE}(\epsilon) =$
> (4) Describe the limit in which both the Fermi-Dirac and Bose-Einstein distribution functions approach the Maxwell-Boltzmann distribution function. Explain your reasoning.
> (5) For a system of identical fermions, using the axes provided, graph the relationship between the distribution function $n_{FD}(\epsilon)$ and energy $\epsilon$ for both $T = 0$ K and $T > 0$ K. Be sure to label the Fermi energy. [Note: Students were provided the axes. Also, in the interviews, students were asked about how the distribution function would change when the temperature changes from one non-zero value to another, e.g., $k_B T \gg E_F$.]

**Figure 1**. Questions students were asked in pre/posttests

These questions map onto the learning goals of the CQS with varying levels of transfer of learning required as shown in Table 1. In the T = 0 K case, question 1 is a near transfer question directly related to CQ10. Question 2, which relates to CQ1 – CQ7, requires further transfer because the system is two dimensional instead of the three dimensional system under consideration in CQ1-CQ7. Note that CQ11 was not administered to students, but would offer greater guidance and support to help students answer pre/posttest questions related to two dimensional free-electron gas. For the T > 0 K case, question 3 is a near transfer from CQ12-CQ15 and the subsequent class discussion, and question 4 is a near transfer from CQ14. Finally, question 5 is not a near transfer of the FD distribution function related concepts in questions CQ12-15 (although they have related concepts), since question 5 in the pre/posttest asks students to use a different representation of knowledge. In particular, students have to use graphical representation to answer question 5. Converting from mathematical representation to graphical representation is not easy for students who are still developing expertise in these topics. We note that direct scaffolding pertaining to graphical representation of the FD distribution function can be provided by instructors after the CQS (this is a suggested topic of class discussion following CQ17). We again emphasize that the learning goals of some clicker questions provided in the appendix are not addressed by these test questions. These clicker questions are provided to give a complete picture of the CQS and offer more of the context in which these topics were being discussed.

**Table 1** _Relation between pre/posttest questions (Q# provided) and various clicker questions (CQS # provided) in the CQS along with comments._

| Q # | CQS # | Comments |
|---|---|---|
| 1 | CQ10 | Near transfer |
| 2 | CQ1-7 | Far transfer, transition from 3D in CQS to 2D in pre/posttests (CQ11 was not administered but could provide additional support for 3D to 2D transition) |
| 3 | CQ12-15 (and class discussion) | Near transfer |
| 4 | CQ14 | Near transfer |
| 5 | CQ12-15 | Far transfer, different representation of knowledge in CQS and pre/posttests (suggested topic for class discussion after CQ17) |

## RESULTS

Student performance on the pretest and posttest provided a measure of student understanding of relevant concepts after traditional instruction and after students engaged with the CQS with peer instruction. Along with the individual interviews, the pretest and posttest also helped us gain an understanding of student difficulties with these concepts. A rubric was developed for grading student performance on the five questions on the pretest and posttest. Two of the authors graded all student responses and the inter-grader reliability was better than 95%. Table 2 compares the in-class pre/posttest performances of students in upper-level undergraduate QM after traditional lecture-based instruction (pretest) and after they had engaged with the CQS on these concepts (posttest). Table 2 also presents the normalized gain ($g$) which is calculated as $g = (post\% - pre\%)/(100\% - pre\%)$ [43]. Moreover, Table 2 displays the effect size on each question between the pre/posttest scores, which was calculated as Cohen's $d = (\mu_{post} - \mu_{pre})/\sigma_{pooled}$ where $\mu_i$ is the mean of group $i$ and the pooled standard deviation is $\sigma_{pooled} = \sqrt{(\sigma_{pre}^2 + \sigma_{post}^2)/2}$ [43]. Table 2 shows that student performance after traditional lecture-based instruction was poor on all questions. After engaging with the CQS, although the average performance improved, it was still at approximately 50% on many of the questions. We note that questions 1 and 4 asked for reasoning for student responses. However, we felt that a student should get some credit for answering a question even if the explanation was not adequately provided (e.g., if a student correctly noted that the Fermi energy is an intrinsic property of the material but did not explain why). Therefore, we graded those questions in two ways, which included or did not include their reasoning. A summary of the student difficulties found in individual interviews and written pre/posttests is presented in Table. 3. We then elaborate on the student difficulties found in interviews and written responses without separating them into pre/posttest since the difficulties were similar after traditional lecture-based instruction and after students engaged with the CQS, although the difficulties were less prevalent after engaging with the CQS (see Table 2).

**Table 2.** *Comparison of the mean pre/posttest scores on each question, normalized gains and effect sizes for students in upper-level undergraduate QM (number of students N=13). The pretest was administered after traditional lecture-based instruction and the posttest after students engaged with the entire CQS on these concepts. The percentages in parentheses for questions 1 and 4 refer to the mean scores when students were not graded for whether the reasoning they provided was correct.*

| # | Pretest Mean | Posttest Mean | Normalized Gain (g) | Effect Size (d) |
|---|---|---|---|---|
| **1a** | 8% (8%) | 46% (46%) | 0.42 | 0.46 |
| **1b** | 50% (62%) | 58% (77%) | 0.15 | 0.09 |
| **2a** | 15% | 77% | 0.73 | 1.91 |
| **2b** | 8% | 54% | 0.50 | 1.23 |
| **2c** | 8% | 50% | 0.46 | 1.33 |
| **3** | 50% | 82% | 0.64 | 0.96 |
| **4** | 29% (33%) | 50% (62%) | 0.29 | 0.45 |
| **5** | 38% | 85% | 0.75 | 1.36 |

**Student difficulties with the Fermi Energy:** In question 1, some students had difficulty with the fact that the Fermi energy of copper is an intrinsic property and provided responses such as the following: "*Cube B has higher Fermi energy because higher states must be filled*", "*The cube with 2N copper atoms because it has a higher free electron density*", "*B has more atoms, thus it encloses a larger surface area in k-spa*ce". One interviewed student, when asked what the Fermi energy is, stated "[It's] *something to do with exclusion principle. An atom with 10 electrons will settle down to the 10 lowest states.* [There's a] *higher Fermi energy with 20 because there are more being pushed up the ladder with higher energy. An additional one will have additional energy.*" This type of reasoning demonstrates either incomplete conceptual understanding of Fermi energy, specifically how the closer level spacing leaves the Fermi energy unchanged. In particular, the difficulty is with the level structure and its dependence on sample size, not the concept of Fermi energy as the energy of the highest occupied level at T = 0 K. It misses the fact that the volume occupied by each state in the k-space has inverse dependence on the volume of the solid, so that the Fermi energy is an intrinsic property of a given material. Some students also had difficulty differentiating between the Fermi energy and total electronic energy and characterized the Fermi energy as the total energy of all the fermions in the system. Responding to question 2a on the posttest in class, none of the students who answered correctly explicitly derived a mathematical expression for the Fermi energy. However, Figure 2 shows the response of a student who answered the conceptual question 1 a incorrectly and attempted to derive an expression for the Fermi energy. Unlike some of the other students who struggled to derive the expression for the Fermi energy in response to question 2 a, Figure 2 shows that this student wrote down correct equations for the Fermi energy while answering question 1 a (which did not explicitly ask for an expression for the Fermi energy). However, the student incorrectly inferred that the Fermi energy depends on the number of copper atoms because he did not take into account the volume in the denominator, i.e., the fact that the Fermi energy depends on the number density of the free electrons. Since the number density (number/volume) of free electrons is constant for copper regardless of the size of the copper cube, the Fermi energy is an intrinsic property of copper. Response to question 2 a in Table 2 shows that a majority of students struggled to derive an expression for the Fermi energy after traditional lecture-based instruction but their performance improved significantly after engaging with the CQS.

**Table 3.** *Summary of conceptual difficulties found (in addition to the quantitative difficulties involving expressions for the Fermi energy, density of states and total electronic energy of a free electron gas). Below each main point, specific examples of difficulties are listed. In comments section, we include the relevant pre/posttest question numbers and whether posttest showed "some" improvement or "major" improvement compared to the pretest.*

| Difficulties | Comments |
|---|---|
| • Difficulty realizing that the Fermi energy is an intrinsic property of copper (difficulty was often related to the level structure and its dependence on sample size, not necessarily directly with the concept of Fermi energy).<br>○ Larger copper cube has higher Fermi energy because it has more free electrons.<br>○ Larger copper cube has higher Fermi energy because more states with higher energy must be filled<br>○ Larger copper cube has more atoms so the cube encloses a larger surface area in k-space (also note the incorrect inference about a larger volume in physical space implying a larger volume in k-space). | 1a and 2a<br>Some improvement |
| • Total electronic energy (TEE) of a free electron gas is the same regardless of the number of atoms<br>○ because it is only dependent on the Fermi energy.<br>○ because it is an intrinsic property of the material.<br>• TEE is lower for the larger copper cube because the degeneracy pressure in the larger system would be lower due to its larger size (and lower degeneracy pressure means lower TEE). | 1b and 2c<br>Some improvement |
| • Difficulties with the density of states were often associated with believing it had something to do with the presence of free electrons (when the presence of particles has nothing to do with it)<br>○ It is the number of free electrons per unit volume in k space.<br>○ It is the number of particles per unit energy.<br>○ It is the number of particles in each single-particle state with energy $\epsilon$.<br>○ The density of particles in a particular energy state.<br>○ In a configuration, how close the occupied states are to one another.<br>○ Particles having energy $\epsilon$ per volume. | 2b<br>Some improvement |
| • Difficulties with Fermi-Dirac, Bose-Einstein and Maxwell-Boltzmann distribution functions<br>○ Not including the chemical potential in the distribution functions.<br>○ Using $e^{\frac{\epsilon - E_F}{k_B T}}$ instead of $e^{\frac{\epsilon - \mu(T)}{k_B T}}$ where $\mu(T)$ is the chemical potential at temperature T and $E_F = \mu(0)$.<br>○ Interchanging the signs $\pm 1$ in the denominator for the fermionic and bosonic cases.<br>○ Stating that occupancy of a single particle state for fermions ranges between ½ and 1.<br>○ Stating that occupancy of a single particle state for fermions is zero or 1 depending on spin. | 3<br>Major improvement |
| • Difficulties with how the Fermi-Dirac and Bose-Einstein distribution functions approach the Maxwell-Boltzmann distribution function (in the high temperature low density limit)<br>○ Stating that a system is in the classical limit in the $T \rightarrow 0\ K$ limit (when, in fact, quantum effects are most pronounced when $T \rightarrow 0\ K$).<br>○ Stating that a system is in the classical limit when single particle energy $\epsilon \rightarrow \infty$ (but one value of $\epsilon$ does not give the entire Fermi-Dirac distribution function, instead, $\epsilon$ is only one single particle energy).<br>○ Stating that a system is in the classical limit when there is no interaction between the particles. | 4<br>Some improvement |
| • Difficulties with the graphical representation of the Fermi-Dirac distribution function<br>○ Stating that at T = 0 K, the Fermi-Dirac distribution function is a delta function at $\epsilon$ = 0 (confusing it with the Bose-Einstein behavior).<br>○ Stating that at T > 0 K, the Fermi-Dirac distribution function as a function of energy is an increasing exponential.<br>○ Stating that at T > 0 K, the Fermi-Dirac distribution function as a function of energy has a peak at a non-zero single particle energy $\epsilon$ and this peak shifts to higher energy $\epsilon$ as the temperature increases (similar to the Maxwell speed distribution for an ideal gas). | 5<br>Major improvement |

$$\frac{1}{8}\left(\frac{4}{3}\pi k_F^3\right) = \frac{N_4}{2}\frac{\pi^3}{V} \rightarrow k_F \propto N^{1/3}$$

$$E_F = \frac{\hbar^2 k_F^2}{2m} \propto N^{2/3}$$

So the cube with 2N copper atoms has higher Fermi energy

**Figure 2**. *A sample response to question 1 a in which the student made an incorrect inference "So the cube with 2N copper atoms has higher Fermi energy" based upon his derivation of a correct expression for the Fermi energy. The student wrote down the equations correctly but did not take into account the volume and the fact that the Fermi energy depends on the number density of the free electrons which is constant for copper independent of the size of the sample.*

**Student difficulties with the density of states**: Some students struggled with the density of states in question 2b with responses such as "$D(\epsilon) = \frac{N}{\frac{\pi k^3}{3}}$" and "$D(\epsilon)=N/\epsilon$". When asked to explain what the density of states means, a common difficulty was relating it to particle number density or probability of occupying a particular single-particle state (confusion between the density of states and the distribution function) as in the following responses: "*This is the number of particles for each energy per k-space volume*", "*number of particles in each single-particle state with energy $\epsilon$*", "*the density of particles in a particular energy state*", "*In a configuration, how close the **occupied** states are to one another*", "*Particles having energy $\epsilon$ per volume*". Moreover, some students appear to have incorrectly interpreted that the density of states has something to do with the number of particles because they calculated the density of states using the expression for the Fermi energy (e.g., see the expression in Fig. 2). In particular, they used an expression similar to that in Fig. 2 to write an expression for the number of particles N in terms of $E_F$ and then calculated $dN/dE_F$ (or $dN/dE_{tot}$ by changing $E_F$ to $E_{tot}$). Although this approach coincidentally gives the correct expression for the density of states, it can mislead students into thinking that density of states is given by the number of particles per unit energy.

**Student difficulties with the total electronic energy**: In response to question 1b, some students stated that the total electronic energy would be the same regardless of the number of copper atoms. Discussions suggest that some of them may have been confused because of the fact that the total electronic energy *per electron* for a free electron gas is $\left(\frac{3}{5}\right) E_F$ and they remembered it incorrectly as $E_{tot} = \left(\frac{3}{5}\right) E_F$. Moreover, in written responses, two students incorrectly claimed that the copper cube with larger *N* will have a lower total electronic energy since the degeneracy pressure in the larger system would be lower due to its larger dimension. The calculation of the total electronic energy in question 2c was extremely difficult for most students because this calculation cannot be done correctly using an algorithmic approach unlike, e.g., the calculations involving traditional circuit problems in introductory physics in which Kirchhoff's rules can be used algorithmically to yield the correct value of current, voltage and resistance in different parts of a complicated circuit without a functional understanding of the underlying concepts. A majority of students struggled to piece together a solution for the total electronic energy. In an interview, before engaging with the CQS, in response to question 2c in the pretest, one student stated, "*I assume I should integrate but I'm not sure how to set it up.*" The following are typical incorrect responses that suggest that different students struggled with different aspects of setting up the integral: "$E = \int_0^{k_F} E_F dk$", "$E_{tot} = \int_0^{k_F} N d\epsilon$", "$E_{tot} = \int D(\epsilon) d\epsilon$", "$E_{tot} = \int_0^{k_F} D(\epsilon) d\epsilon$", "$E = \int_0^{k_{max}} \frac{\hbar^2 k^2}{2m} dk = \frac{\hbar^2 k_{max}^3}{6m}$", "$E = \int_0^{k_F} \frac{\hbar^2 k^2}{2m} \sigma l_x l_y dk = \frac{\hbar^2 k_{max}^3}{6m}$, $l_x$ is length of square in x, $l_y$ is length of square in y, $\sigma$ is number of free electrons per unit area". Table 2 shows that the average student scores on questions 2b and 2c related to the calculation of the density of states and total electronic energy improved from less than 10% after traditional lecture-based instruction to approximately 50% after the CQS. Deriving these expressions was extremely challenging for many students despite the fact that the only difference between the CQS and the pre/posttest questions 2b and 2c is that the dimensionality of the system was two dimensions, rather than three dimensions. Some interviewed students needed guidance from the interviewer to successfully calculate the total electronic energy in question 2c even on the posttest. Other interviewed students (not included in Table 2) asked to review CQ1-CQ7 again, which were in the three dimensional context, before answering question 2c in two dimensions on the posttest.

**Student difficulties with the expressions for the distribution functions and constraints on the number of particles in each single-particle state:** Table 2 shows that out of all of the questions, students performed relatively well on both pre/posttests on question 3 which asked for the expressions for the distribution functions and constraints on the number of particles in each single-particle state. On the pretest, average student scores were 39% on the distribution functions and 61% on the constraints and on the posttest, average student scores were 85% on the distribution functions and 80% on the constraints. With regard to the distribution function, some students did not include the chemical potential in the expressions or interchanged the signs in the denominator for the fermionic and bosonic cases. Also, although we did not take off points for students who wrote $e^{(\epsilon-E_F)/k_B T}$ instead of $e^{(\epsilon-\mu(T))/k_B T}$, where $\mu(T)$ is the chemical potential at temperature $T$ and $E_F = \mu(0)$, in the expression for the Fermi-Dirac distribution function, the temperature dependence of $\mu$ is crucial for the low-temperature specific heat of metals and other effects such as thermoelectricity. The most common difficulty for the constraints was swapping the fermionic and bosonic cases. Other incorrect responses include claims that the constraint on the number of particles in each single particle state for the fermionic case is between ½ and 1 particle or that it is either 0 or 1 depending on spin.

**Student difficulties with the high temperature limiting case of quantum distribution functions**: Students had difficulty with the condition under which the Fermi-Dirac and Bose-Einstein distribution functions approach the classical limit of the Maxwell-Boltzmann distribution function. In the high temperature limit $T \to \infty$, the de Broglie wavelength $\lambda$ of the particle wave becomes very small and the overlap of the wavefunctions of different particles in the system becomes negligible (i.e., $\lambda \left(\frac{N}{V}\right)^3 \ll 1$ where $N/V$ is the number of free electrons per unit volume). In this limit, the Fermi-Dirac and Bose-Einstein distribution functions approach the Maxwell-Boltzmann distribution function. Students had great difficulty with question 4 which focused on this issue (see Table 2). The student difficulties on this question can be classified in a few categories as follows.

Some students claimed that the quantum distribution functions will approach the MB distribution function when $T \to 0\ K$ (which is the exact opposite case in which the quantum effects are important). These students often focused on the mathematical expressions for the quantum distribution functions and mathematically reasoned about how they might approach the MB distribution function. Interviews suggest that students with these types of responses who resorted to using mathematical expressions as the basis for their answer often did not think physically about whether their mathematical reasoning made sense conceptually. This dichotomy of either being in the "math" mode (which was prevalent for students who claimed the correct limit was $T \to 0\ K$) or the "physics" mode and not integrating the mathematical and physical reasoning to do the sense-making is a common novice-like problem solving approach and has been observed in other contexts [23,36]. Two typical responses in this category are: "*When $T \to 0$, $\frac{1}{e^{\frac{\epsilon-\mu}{k_B T}}\pm 1} \sim \frac{1}{e^{\frac{\epsilon-\mu}{k_B T}}}$*", and "*They approach Maxwell-Boltzmann distribution when the exponential part is the most important, so when $T \to 0$*". Figure 3 shows another student response that falls in this category. Figure 3 shows that the student writes the correct distribution functions in response to question 3 and states in response to question 4 that "*When the $e^{\frac{\epsilon-\mu}{k_B T}}$ term is much larger than 1, the distributions are roughly the same. This happens when $\epsilon > \mu$ and low temperatures $T \to 0$.*" This response is interesting because the student explicitly notes that the exponential term $e^{\frac{\epsilon-\mu}{k_B T}}$ is much larger than 1 when $\epsilon > \mu$ and $T \to 0$ K which is correct but did not contemplate the case when $\epsilon < \mu$ and $T \to 0\ K$. For a fermionic system, this latter case ($\epsilon < \mu$ and $T \to 0\ K$) yields $n_{FD}(\epsilon) = 1$ for all states below the Fermi energy (Fermi energy is the chemical potential at $T = 0$ K). In other words, in the limit $T \to 0$ K, the system does not behave classically. In fact, this limit $T \to 0$ K that the student states is the classical limit is the most quantum mechanical case for when the Fermi Dirac distribution function is a step function with all states below the Fermi energy completely filled and all states above the Fermi energy completely empty. The student did not consider the states below the Fermi energy which if analyzed correctly would show that all of those states are filled and the Fermi-Dirac distribution function in this case of $T \to 0$ K is a step function. This type of response suggests that focusing only on mathematical reasoning for part of the problem correctly (i.e., $\epsilon > \mu$) but not reflecting on the conceptual aspect of what should happen in the $T \to 0$ K limit may have at least partly prevented the student from realizing that he did not take into account the $\epsilon < \mu$ and $T \to 0$ K situation. Interviews also suggest that some of the students had inadequate understanding of the chemical potential of the system which exacerbated the difficulty in reasoning about the limiting case. For example, students did not realize that $\mu$ for particles with mass depends on temperature and for fermions, $\mu$ is equal to the Fermi energy at $T = 0$ K and then it decreases with an increase in temperature and eventually becomes negative at high temperatures (while for bosons, $\mu$ is zero at and below the critical temperature and is negative otherwise).

$$FD: \frac{1}{e^{(\epsilon-\mu)/k_BT}+1} \qquad B\text{-}E: \frac{1}{e^{(\epsilon-\mu)/k_BT}-1} \qquad MB: \frac{1}{e^{(\epsilon-\mu)/k_BT}}$$

When the $e^{(\epsilon-\mu)/k_BT}$ term is much larger than 1, the distributions are roughly the same. This happens when $\epsilon > \mu$ and low temperatures, $T \to 0$.

**Figure 3**. *A sample response in which the student wrote the correct mathematical expressions for each of the distributions functions but drew an incorrect conclusion about the limiting case.*

Other students who mainly reasoned using the expressions for the distribution functions focused only on large energy $\epsilon$ and associated the classical limit with $\epsilon \to \infty$. They claimed that the quantum distribution functions will approach the MB distribution function when $\epsilon \to \infty$, as in the following student responses: "*At very large $\epsilon$, $\frac{\epsilon-\mu}{k_BT} \gg 1$ so it approaches M-B distribution*", "*As $\epsilon \to \infty$, all converge b/c the $\pm 1$ in denominator becomes irrelevant*", "*In the high $\epsilon$-limit, as the exponential term becomes very large*".

Some students claimed that the quantum distribution functions will approach the MB distribution function whenever the particles are non-interacting, as in the following responses: "*If there is no interaction, both of them* [Fermi-Dirac and Bose-Einstein] *become classical* [Maxwell-Boltzmann]" or "*If there is no interaction…classical.*" Discussions suggest that these students often confused the overlap of the wavefunction of different particles in the system being negligible (which happens in the high temperature limit) with the particles being non-interacting or Coulomb interaction between the electrons being negligible to approach the MB distribution limit.

**Student difficulties with the graphical representation of the Fermi-Dirac distribution function**: The ability to transform from one representation of knowledge to another, e.g., mathematical to graphical, is a sign of expertise. Experts often transform from one representation of knowledge to another to simplify the problem solving process. Table 2 shows that on question 5 on the pre/posttests that asked students to draw the Fermi-Dirac distribution function at *T=0 K* and *T > 0 K* students' average score more than doubled after engaging with the CQS following traditional lecture-based instruction. On the pretest, students struggled with the graphical representation of the Fermi-Dirac distribution function both at zero and non-zero temperatures. For example, Figure 4 shows one such graph on which an interviewed student drew an exponentially increasing $n_{FD}(\epsilon)$ vs. $\epsilon$ at *T > 0 K*.

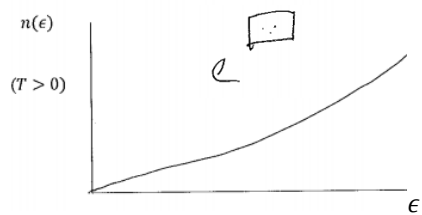

**Figure 4**. *An interviewed student's incorrect graphical representation of the Fermi-Dirac distribution function at T > 0K stating the function has exponential shape because the expression for $n_{FD}(\epsilon)$ involves "…e to the something… probably has energy in there. I'm not so sure just how temperature would fit into it though."*

Another interviewed student incorrectly claimed that as the temperature increases, a peak appears in $n_{FD}(\epsilon)$ and that peak in the Fermi-Dirac distribution function would shift to higher temperatures and the occupation of the ground state would eventually reach zero (see his drawing in Figure 5). "*For T >0, there's a local maximum at $\epsilon$=0, but it gets pushed out as T increases. Much, much greater, and it would get pushed out past the Fermi energy. I wouldn't think there should be anything keeping that maximum at the Fermi energy.*" Discussion suggests that the student may have been confused about the peak in the Maxwell speed distribution (which has the square of the speed from the volume element multiplying the exponential factor so that there is a peak at a non-zero value of speed) and how the peak in that distribution moves to higher energies as the temperature increases. In written responses to question 5 also, some students drew the Fermi-Dirac distribution function with peaks at the Fermi energy or some other non-zero energy similar to Figure 5. For example, one student who drew the Fermi-Dirac distribution function correctly at *T=0 K*, drew a graph similar to that shown in Figure 5 for *T >>0 K*. Another student drew a graph similar to that shown in Figure 5 for *T >>0 K* for both *T=0 K* and *T > 0 K* with peaks of different heights centered at the same value of single-particle energy. Two students who confused the fermionic and bosonic distribution functions, drew the Fermi-Dirac distribution function to be a delta function at $\epsilon = 0$ at *T = 0 K*.

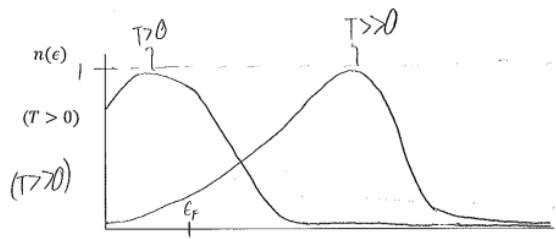

**Figure 5**. *An interviewed student's incorrect graphical representation of the Fermi-Dirac distribution function in which the student incorrectly stated that the single-particle ground state of the fermions would eventually be vacated and the peak in the distribution function (which is not supposed to be there) will keep shifting to higher energies as the temperature of the system increased.*

## DISCUSSION, SUMMARY AND FUTURE PLANS

We investigated the difficulties that physics students in upper-level undergraduate QM and graduate students after quantum and statistical mechanics core courses have with the Fermi energy, Fermi-Dirac distribution and total electronic energy of a free electron gas after they had learned relevant concepts in their respective courses. These difficulties were probed by administering to undergraduate students written conceptual questions and questions involving understanding of the procedures (to derive various expressions) and asking students in undergraduate and graduate courses to answer those questions while thinking aloud [35] in one-on-one individual interviews. We find that advanced students have many common difficulties with these concepts, after traditional lecture-based instruction and students struggled with both conceptual and quantitative questions.

We also investigated the impact of a validated CQS on these topics on undergraduate student performance. The CQS questions combined conceptual and quantitative knowledge because the CQS strived to help students learn the relevant concepts as well as the ability to solve problems, e.g., derive expressions for the Fermi energy, density of states and total electronic energy of a free electron gas. It was unclear a-priori how much these advanced students will benefit from the CQS that integrates conceptual and quantitative knowledge. The implementation of the CQS in an upper-level undergraduate QM course shows that while engaging with the CQS reduced these difficulties, many advanced students continued to struggle with these challenging concepts. In particular, student performance in the undergraduate course improved after students engaged with the CQS on relevant concepts but there is still substantial room for improvement. The improvement from the pretest to posttest sheds light on the expertise of advanced students and their ability to learn from the CQS that combined conceptual and quantitative knowledge. Since the pre/posttest questions are open-ended, students must transfer their learning from the CQS and generate the responses to the pre/posttest questions without any scaffolding support. Although the normalized gains and effect sizes in Table 2 are noteworthy partly because students performed poorly on the pretest after traditional instruction, the actual posttest scores are not impressive. It appears that combining the conceptual and quantitative knowledge in the CQS was not sufficient to help advanced students develop a solid grasp of these concepts and understanding of the procedures in order to be able to generate them without support in posttest.

The future refinements of the CQS will focus on incorporating more student difficulties that were found. In particular, the current version of the CQS incorporates only some of the difficulties found, e.g., the difficulty associated with the volume of a shell in k space in CQ3, relating the density of states to the number of particles in CQ5, incorrect relation between the density of states and the total electronic energy in CQ6, believing that Fermi energy is an extensive quantity in CQ10, etc. However, incorporating more student difficulties may increase the effectiveness of the CQS further. For example, in order to better address the difficulties with the distribution function, an increased emphasis on the chemical potential, its role in the distribution functions, and its behavior as a function of temperature can be included in the CQS via additional clicker questions. The difficulties with the limiting case posed in question 4 can also be addressed more explicitly via additional clicker questions. Moreover, although we hypothesized that students should be guided through problems involving understanding of the procedures (i.e., problems involving both conceptual and quantitative parts) via a series of clicker questions as in CQ1-CQ7, student learning may be improved for question 2 which involves calculations of the Fermi energy, density of states and total electronic energy by changing the implementation of the CQS. In fact, a majority of interviewed students required at least some guidance from the interviewer via leading questions in order to do the calculations correctly in question 2 on the posttest, which focused on a two dimensional system instead of the three dimensional system treated in CQ1-CQ7. Some interviewees asked to review CQ1-CQ7 one more time for the three dimensional calculations before calculating the corresponding quantities in two dimensions in question 2 on the posttest. In the future implementation of the CQS, immediately after the students engage with clicker questions CQ1-CQ7 in the class, we plan to ask them to respond to question 2 without any support so that they have an opportunity to reflect upon their proficiency in deriving the expressions for the Fermi energy, density of states and total electronic energy of a free electron gas and tell them right before they engage with the CQS that they will have to do the derivations immediately after the CQS as a quiz. This type of immediate individual reflection after engaging with the CQS with their peers may be helpful in improving individual accountability and focus, and help students consolidate the concepts learned and develop facility and self-reliance in calculating these quantities by combining conceptual and quantitative problem solving.

Moreover, future studies would look at the impact of developing, validating and implementing a guided tutorial worksheet related to these topics in which students are guided through these issues but are not provided the level of scaffolding that is provided via the clicker questions. In particular, in answering the CQS, students have to only identify the correct choices instead of generating them as would be the case when working through a tutorial. Thus, the tutorial approach, which requires students to generate the answers, may provide more productive struggle than the CQS approach and has the potential to help students develop a better grasp of these challenging concepts and understand ing of the procedures. The findings discussed here can be useful in the development of such a tutorial worksheet.

## ACKNOWLEDGEMENT
We thank the National Science Foundation for award PHY-1806691 and Prof. R. P. Devaty for helpful discussions.

## REFERENCES

1. Jolly P, Zollman D, Rebello S and Dimitrova A 1998 Visualizing potential energy diagrams *Am J Phys* **66** 57
2. Ireson G 1999 A multivariate analysis of undergraduate physics students' conceptions of quantum phenomena *Eur J Phys* **20** 193
3. Zollman D, Rebello N, and Hogg K 2002 Quantum mechanics for everyone: Hands-on activities integrated with technology *Am J Phys* **70** 252
4. Muller R and Wiesner H 2002 Teaching quantum mechanics on an introductory level *Am J Phys* **70** 200
5. Wittmann M et al. 2002 Investigating student understanding of quantum physics: Spontaneous models of conductivity *Am J Phys* **70** 218
6. Domert D, Linder C and Ingerman A 2005 Probability as a conceptual hurdle to understanding one-dimensional quantum scattering and tunneling *Eur J Phys* **26** 47
7. Kohnle A et al. 2010 Developing and evaluating animations for teaching quantum mechanics concepts *Eur J Phys* **31** 1441
8. Singh C 2001 Student understanding of quantum mechanics *Am J Phys* **69** 885; Singh C 2005 Transfer of learning in quantum mechanics, AIP Conf Proc Melville NY **790**, 23 https://doi.org/10.1063/1.2084692; Singh C 2006 Assessing and improving student understanding of quantum mechanics, AIP Conf Proc Melville NY **818**, 69 https://doi.org/10.1063/1.2177025
9. Marshman E and Singh C 2019 Validation and administration of a conceptual survey on the formalism and postulates of quantum mechanics *Phys Rev PER* **15** 020128
10. Kohnle A et al. 2014 A new introductory quantum mechanics curriculum *Eur J Phys* **35** 015001
11. Gire E et al. 2015 The structural features of algebraic quantum notations *Phys Rev ST PER* **11** 020109
12. Dini V and Hammer D 2017 Case study of a successful learner's epistemological framings of quantum mechanics *Phys Rev PER* **13**, 010124
13. Passante G et al. 2015 Examining student ideas about energy measurements on quantum states across undergraduate and graduate level *Phys Rev ST PER* **11**, 020111.
14. Johansson A 2018 Undergraduate quantum mechanics: lost opportunities for engaging motivated students? *Eur J Phys* **39** 025705
15. Singh C, Belloni M and Christian W 2006 Improving student's understanding of quantum mechanics, *Phys Today* **59** 43; Singh C 2008 Interactive learning tutorials on quantum mechanics *Am J Phys* **76** 400
16. Singh C and Zhu G 2009 Cognitive issues in learning advanced physics: An example from quantum mechanics in *Proc Phys Educ Res Conf,* AIP Conf Proc Melville NY **1179** p. 63 https://doi.org/10.1063/1.3266755; Singh C 2008 Student understanding of quantum mechanics at the beginning of graduate instruction *Am J Phys* **76** 277
17. Lin S, and Singh C 2010 Categorization of quantum mechanics problems by professors and students *Eur J Phys* **31** 57
18. Zhu G and Singh C 2011 Improving students' understanding of quantum mechanics via the Stern-Gerlach experiment, Am. J. Phys. **79**, 499; Zhu G and Singh C 2009 Students' understanding of Stern Gerlach experiment, *Proc Phys Educ Res Conf,* AIP Conf Proc Melville NY **1179** p. 309 https://doi.org/10.1063/1.3266744; Justice P and Singh C 2019 Improving student understanding of quantum mechanics underlying the Stern-Gerlach experiment using a research-validated multiple-choice question sequence *Eur J Phys* **40** 055702

19. Zhu G and Singh C 2012 Surveying students' understanding of quantum mechanics in one spatial dimension *Am J Phys* **80** 252; Zhu G and Singh C 2012 Improving students' understanding of quantum measurement: I. Investigation of difficulties *Phys Rev ST PER* **8** 010117; Zhu G and Singh C 2012 Improving students' understanding of quantum measurement: II. Development of research-based learning tools *Phys Rev ST PER* **8** 010118
20. Zhu G and Singh C 2013 Improving student understanding of addition of angular momentum in quantum mechanics, *Phys Rev ST PER* **9**, 010101; Singh C 2007 Student difficulties with quantum mechanics formalism in *Proc Phys Educ Res Conf,* AIP Conf Proc Melville NY **883,** p. 185 https://doi.org/10.1063/1.2508723
21. DeVore S and Singh C 2015 Development of an interactive tutorial on quantum key distribution *Proc Phys Educ Res Conf* https://doi.org/10.1119/perc.2014.pr.011; Singh C 2007 Helping students learn quantum mechanics for quantum computing in *Proc Phys Educ Res Conf* , AIP Conf Proc Melville NY **883,** p. 42 https://doi.org/10.1063/1.2508687
22. Brown B and Singh C 2015 Development and evaluation of a quantum interactive learning tutorial on Larmor precession of spin, *Proc Phys Educ Res Conf* https://doi.org/10.1119/perc.2014.pr.008
23. Singh C and Marshman E 2015 Review of student difficulties in quantum mechanics *Phys Rev ST PER* **11** 020119; Marshman E and Singh C 2015 Framework for understanding student difficulties in quantum mechanics *Phys Rev ST Phys Educ Res* **11** 020117; Mason A and Singh C 2010 Do advanced students learn from their mistakes without explicit intervention *Am J Phys* **78** 760; Brown B, Mason A and Singh C 2016 Improving performance in quantum mechanics with explicit incentives to correct mistakes *Phys Rev ST PER* **12** 010121
24. Marshman E and Singh C 2016 Interactive tutorial to improve student understanding of single photon experiments involving a Mach-Zehnder interferometer *Eur J Phys* **37** 024001
25. Siddiqui S and Singh C 2017 How diverse are physics instructors' attitudes and approaches to teaching undergraduate-level quantum mechanics? *Eur J Phys* **38,** 035703
26. Marshman E and Singh C 2017 Investigating and improving student understanding of quantum mechanics in the context of single photon interference, *Phys Rev PER* **13**, 010117
27. Marshman E and Singh C 2017 Investigating and improving student understanding of quantum mechanical observables and their corresponding operators in Dirac notation *Eur J Phys* **39** 015707
28. Marshman E and Singh C 2017 Investigating and improving student understanding of the expectation values of observables in quantum mechanics *Eur J Phys* **38** (4) 045701
29. Marshman E and Singh C 2017 Investigating and improving student understanding of the probability distributions for measuring physical observables in quantum mechanics *Eur J Phys* **38** (2) 025705
30. Maries A, Sayer R and Singh C 2017 Effectiveness of interactive tutorials in promoting "which-path" information reasoning in advanced quantum mechanics *Phys Rev PER* **13**, 020115
31. Sayer R, Maries A and Singh C 2017 A quantum interactive learning tutorial on the double-slit experiment to improve student understanding of quantum mechanics *Phys Rev PER* **13,** 010123
32. Keebaugh C, Marshman E and Singh C 2018 Improving student understanding of corrections to the energy spectrum of the hydrogen atom for the Zeeman effect *Phys Rev PER* **15** 010113; Keebaugh C, Marshman E and Singh C 2018 Investigating and addressing student difficulties with the corrections to the energies of the hydrogen atom for the strong and weak field Zeeman effect *Eur J Phys* **39**, 045701; Keebaugh C, Marshman E and Singh C 2019 Improving student understanding of fine structure corrections to the energy spectrum of the hydrogen atom *Am J Phys* **87** 594
33. Keebaugh C, Marshman E and Singh C 2018 Investigating and addressing student difficulties with a good basis for finding perturbative corrections in the context of degenerate perturbation theory *Eur J Phys* **39**, 05570; Keebaugh C, Marshman E and Singh C 2019 Improving student understanding of a system of identical particles with a fixed total energy *Am J Phys* **87** 583
34. Sharma S and Ahluwalia P 2012 Diagnosing alternative conceptions of Fermi energy among undergraduate students *Eur J Phys* **33** 883; Griffiths D, *Introduction to Quantum Mechanics* 2nd Ed (Prentice Hall, Upper Saddle River, NJ, 1995)
35. Chi M, Thinking aloud, in *The Think Aloud Method: A Practical Guide to Modeling Cognitive Processes*, edited by Van Someren M, Barnard Y, and Sandberg J (Academic Press, London, 1994)
36. Mazur E, *Peer Instruction: A User's Manual* (Prentice Hall, Upper Saddle River, NJ, 1997)
37. Meltzer D and Manivannan K 2002 Transforming the lecture-hall environment: The fully interactive physics lecture *Am J Phys* **70** 639
38. Novak G, Patterson E, Gavrin A, Christian W and Forinash K 1999 Just in time teaching, *Am J Phys* **67**, 937
39. Sayer R, Marshman E, and Singh C 2016 A case study evaluating Just-in-Time Teaching and Peer Instruction using clickers in a quantum mechanics course *Phys Rev PER* **12** 020133
40. Singh C and Zhu G 2012 Improving students' understanding of quantum mechanics by using peer instruction tools in *Proc Phys Educ Res Conf*, AIP Conf. Proc., Melville, New York **1353**, 77 https://doi.org/10.1063/1.3679998

41. Ding L et al. 2009 Are we asking the right questions? Validating clicker question sequences by student interviews *Am J Phys* **77** 643
42. Heller P, Keith R and Anderson S 1992 Teaching problem solving through cooperative grouping. 1. Group vs individual problem solving *Am J Phys* **60** 627
43. Cohen J 1988 *Statistical Power Analysis for the Behavioral Sciences* (L. Erlbaum Associates)

# APPENDIX: CLICKER QUESTION SEQUENCES

*Notation and procedures coincide with treatment by Griffiths (2<sup>nd</sup> Edition). Correct answers are **bolded**.*

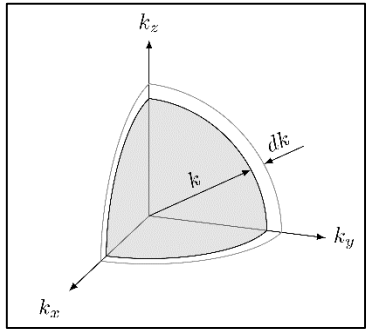

**Figure 6**. *Figure for CQ1-CQ10.*

### Section 1 (T = 0 K):

**(CQ1)** *Let's consider an octant in k-space. Choose all of the following statements that are correct about the k-space for a free electron gas at T = 0 K in a large cubical box of volume V made up of N atoms each with q free electrons, given that two electrons with opposite spins occupy a volume $\frac{\pi^3}{V}$ in k-space.*

(1) *In k-space, the volume of an octant with highest occupied wave vector $k_F$ is*
$\frac{1}{8}(volume\ of\ sphere\ with\ radius\ k_F) = \frac{1}{8}\left(\frac{4}{3}\pi k_F^3\right) = \left(\frac{1}{6}\pi k_F^3\right).$
(2) *The total volume occupied by the free electrons is*
$\frac{(Total\ number\ of\ free\ electrons)}{2}$ *(Volume occupied by two electrons in $k$ − space with opposite spins)* $= \frac{Nq}{2}\left(\frac{\pi^3}{V}\right).$
(3) *The volume of an octant with highest occupied wave vector $k_F$ and the total volume occupied by the free electrons in k-space are equal, so* $\left(\frac{1}{6}\pi k_F^3\right) = \frac{Nq}{2}\left(\frac{\pi^3}{V}\right).$

A. *1 only*    B. *2 only*    C. *1 and 2 only*    D. *2 and 3 only*    **E**. *all of the above*

**(CQ2)** *Equating $\left(\frac{1}{6}\pi k_F^3\right) = \frac{Nq}{2}\left(\frac{\pi^3}{V}\right)$ and solving for $k_F$ gives $k_F = (3\pi^2\rho)^{1/3}$ where $\rho = \frac{Nq}{V}$ is the free electron density (number of free electrons per unit volume). Using this information, choose all of the following statements that are correct about the Fermi energy $E_F$ for non-interacting electrons in the free electron gas model.*

1) *The Fermi energy $E_F$ is the energy of the highest occupied state at $T = 0\ K$.*
2) *The Fermi energy is $E_F = \frac{\hbar^2 k_F^2}{2m} = \frac{\hbar^2}{2m}(3\pi^2\rho)^{2/3}$*
3) *The Fermi energy $E_F$ only depends on the electron number density and the mass of the electron.*

A. *1 only*    B. *1 and 2 only*    C. *1 and 3 only*    D. *2 and 3 only*    **E**. *all of the above*

**(CQ3)** *Choose all of the following statements that are correct about a free electron gas in three dimensions.*
1) *Each state in each shell between k and k+dk has energy $\epsilon = \frac{\hbar^2 k^2}{2m}$.*
2) *The volume of a shell of thickness $dk$ between k and k+dk in the relevant octant in k space occupied by free electrons*

is $\frac{1}{8}\left(\frac{4}{3}\pi k^3\right)dk$.
3) The volume of a shell of thickness dk between k and k+dk in the relevant octant in k space occupied by free electrons is $\frac{1}{8}(4\pi k^2)dk$.

A. 2 only    B. 3 only    C. 1 and 2 only    **D**. 1 and 3 only    E. None of the above.

---

*(CQ4)* Choose all of the following statements that are correct for the 3D free electron gas model. (Include electron spin when relevant.)

1) The number of electron states in each shell between k and k+dk is:

$$\frac{2 \times Volume\ of\ Shell}{Volume\ associated\ with\ a\ single\ state\ in\ k-space} = \frac{2 \cdot \frac{1}{8}(4\pi k^2)dk}{\frac{\pi^3}{V}} = \frac{V}{\pi^2}k^2 dk.$$

2) The total energy of the electrons in a shell between k and k + dk is $\frac{\hbar^2 k^2}{2m}\frac{V}{\pi^2}k^2 dk$ where $\epsilon = \frac{\hbar^2 k^2}{2m}$.

3) The total electronic energy of the system at T = 0 K can be calculated as:
$$E_{tot} = \int_0^{k_F} \frac{\hbar^2 k^2}{2m}\frac{V}{\pi^2}k^2 dk = \frac{\hbar^2}{2m}\frac{V}{\pi^2}\int_0^{k_F} k^4 dk = \frac{\hbar^2}{2m}\frac{V}{\pi^2}\frac{k_F^5}{5}$$

A. 1 only    B. 1 and 2 only    C. 1 and 3 only    D. 2 and 3 only    **E**. All of the above.

---

*(CQ5)* Choose all of the following statements that are correct about a free electron gas <u>in three dimensions</u>. (Include electron spin when relevant. $\epsilon$ defines a surface in k-space. A shell is the volume between two closely-spaced energy surfaces.)

1) The number of states in each shell between $\epsilon$ and $\epsilon+d\epsilon$ is $D(\epsilon)\, d\epsilon$ where the density of states, $D(\epsilon)$, is the number of states per small interval of energy for a given $\epsilon$.
2) The density of states, $D(\epsilon)$, is the number of particles in a given energy interval between energy $\epsilon$ and $\epsilon+d\epsilon$.
3) For a free electron gas at T = 0 K, the total energy, $E_{tot}$, can be calculated as $E_{tot} = \int_0^{E_F} \epsilon\, D(\epsilon) d\epsilon$.

A. 1 only    B. 1 and 2 only    **C**. 1 and 3 only    D. 2 and 3 only    E. All of the above.

---

*(CQ6)* Consider the following conversation:

Student 1: "For a free electron gas at T = 0 K, the total electronic energy $E_{tot} = \int_0^{E_F} D(\epsilon)d\epsilon$."
Student 2: "I disagree. Your integral gives the total <u>number of states</u> up to the Fermi energy."
Student 3: "The total energy is $E_{tot} = \int_0^{E_F} \epsilon\, D(\epsilon)d\epsilon$, which can also be calculated in terms of k as we did in a preceding question."

A. Student 1 only    B. Student 2 only    C. Students 1 and 3 only    **D**. Students 2 and 3 only    E. All of the above.

---

*(CQ7)* Choose all of the following statements that are correct about the density of states, $D(\epsilon)$, for the **3D free electron gas model** given that the number of electron states in the shell between k and k+dk is $\frac{V}{\pi^2}k^2 dk$ (electron spin has been included).

1) $D(\epsilon)d\epsilon = \frac{V}{\pi^2}k^2 dk$. Therefore, $D(\epsilon)d\epsilon$ is proportional to $k^2 dk$.
2) Given that $\epsilon = \frac{\hbar^2 k^2}{2m}$, $d\epsilon = \frac{\hbar^2}{m}k dk$.
3) Using (1) and (2), $D(\epsilon)d\epsilon$ is proportional to $\sqrt{\epsilon}d\epsilon$. Therefore $D(\epsilon)$ is proportional to $\sqrt{\epsilon}$.

A. 1 only    B. 1 and 2 only    C. 1 and 3 only    D. 2 and 3 only    **E**. All of the above.

**(CQ8)** We now know that the total electronic energy of the system of free electrons at $T = 0K$ is $E_{tot} = \left(\frac{\hbar^2(3\pi^2 Nq)^{5/3}}{10\pi^2 m}\right) V^{-2/3}$, which can be written as a function of volume given that the number of free electrons is fixed: $E_{tot} = C V^{-2/3}$. Then $E_{tot} = (-2/3)C V^{-5/3} dV$. Choose all of the following that are correct. (Assume no heat transfer from the free electron gas to its surroundings in the process discussed below).

1) If $dV$ is negative, $dE_{tot}$ is positive.
2) The infinitesimal work by the system $dW = PdV = -dE_{tot}$, so the "degeneracy pressure" $p = (\frac{2}{3})C V^{-5/3}$.
3) The degeneracy pressure is due to the anti-symmetrization requirement of the many-particle wavefunction of the electrons.
4) Since decreasing the volume of the system increases the energy, it is not energetically favorable.

A. 1 and 2 only   B. 1 and 3 only   C. 2 and 3 only   D. 1, 2, and 3 only   E. All of the above.

---

**(CQ9)** Choose all of the following statements that are correct about a free electron gas system.

1) "Degeneracy pressure" describes the quantum mechanical effect that there is an "outward" force per unit area that prevents a solid from collapsing due to the restriction that each electron must occupy a different single particle state.
2) The degeneracy pressure plays a role in stabilizing a solid object.
3) If the electrons were bosons, the total electronic energy of the free electron gas at temperature $T = 0K$ would be lower than what it actually is.

A. 1 only   B. 3 only   C. 1 and 2 only   D. 1 and 3 only   E. All of the above.

---

**(CQ10)** Cubes A and B with the same atom number density have N and 2N sodium atoms, respectively. Choose all of the following statements that are correct.

1) At temperature $T = 0$ K, the Fermi energy of sodium in cube B is larger than the Fermi energy of sodium in cube A.
2) At temperature $T = 0$ K, the total electronic energy of the electrons in cube B is larger than the total electronic energy of the electrons in cube A.
3) If we slowly compress the volume of cube A, the total electronic energy of the electrons in cube A will increase.

A. 1 only   B. 2 only   C. 1 and 2 only   **D. 2 and 3 only**   E. All of the above.

---

**Class Discussion (3D free electron gas at T = 0 K)**

- Two electrons in the same spatial state with "opposing" spins occupy a volume $\frac{\pi^3}{V}$ in k-space (regardless of the value of $\vec{k}$).
- The volume $\frac{\pi^3}{V}$ in k-space can accommodate two electrons due to the fact that electrons are spin-1/2 fermions.
- For an octant, the volume of a shell between $k$ and $k+dk$ of thickness $dk$ is $\frac{1}{8}(4\pi k^2)dk$.
- The number of electron states in the shell between $k$ and $k+dk$ is $\frac{V}{\pi^2} k^2 dk$, which is equal to $D(\epsilon)d\epsilon$ where $D(\epsilon)$ is the density of states with $\epsilon = \frac{\hbar^2 k^2}{2m}$.
- The energy of the electrons in a shell of thickness $dk$ is $dE = \frac{\hbar^2 k^2}{2m} \frac{V}{\pi^2} k^2 dk$.
- The total electronic energy is $E_{tot} = \frac{\hbar^2}{2m} \frac{V}{\pi^2} \int_0^{k_F} k^4 dk = \frac{\hbar^2}{2m} \frac{V}{\pi^2} \frac{k_F^5}{5} = \left(\frac{\hbar^2(3\pi^2 Nq)^{5/3}}{10\pi^2 m}\right) V^{-2/3}$.
- "Degeneracy pressure" is an "outward" quantum mechanical force per unit area that prevents the system from collapsing.

*The following question was not administered in this study but could help scaffold transfer from 3D to 2D in future implementations.*

**(CQ11)** Choose all of the following statements that are correct about the density of states, $D(\epsilon)$, for the **2D free electron gas model** given that the number of electron states in the shell between k and k+dk is $\frac{A}{\pi}kdk$ (electron spin has been included).

1) $D(\epsilon)d\epsilon = \frac{A}{\pi}kdk$. Therefore, $D(\epsilon)d\epsilon$ is proportional to $kdk$.
2) Given that $\epsilon = \frac{\hbar^2 k^2}{2m}$, $d\epsilon = \frac{\hbar^2}{m}kdk$.
3) Using (1) and (2), $D(\epsilon)d\epsilon$ is proportional to $d\epsilon$, therefore $D(\epsilon)$ does not depend on energy $\epsilon$.

A. 1 only    B. 1 and 2 only    C. 1 and 3 only    D. 2 and 3 only    **E. All of the above.**

---

*Section 2 (System at T > 0 K):*

**(CQ12)**

- The distribution function $n(\epsilon)$ is the average number of particles in a given single particle state with energy $\epsilon$.
- For a given system, $D(\epsilon)$, the **density of states** for energy $\epsilon$, is the number of single particle states per unit energy with energy $\epsilon$.
- The average number of particles per unit energy with energy $\epsilon$ is $N(\epsilon)$.

Choose all of the following statements that are correct:

1) $n(\epsilon) = \frac{N(\epsilon)}{D(\epsilon)}$
2) Fermions, bosons, and distinguishable particles have different $n(\epsilon)$.
3) Bosons and distinguishable particles have the same $n(\epsilon)$ because there is no limit to the number of particles that can occupy a given single-particle state, unlike fermions.

A. 1 only    B. 2 only    C. 3 only    **D. 1 and 2 only**    E. 1 and 3 only.

---

**(CQ13)** Choose all of the following statements that are correct about $n(\epsilon)$ (all symbols have their usual meaning):

1) The distribution function $n(\epsilon)$ is the average number of particles in a given single-particle state with energy $\epsilon$.
2) The distribution function is $n(\epsilon) = e^{\frac{-(\epsilon-\mu)}{k_B T}}$ in situations in which the particles can be treated as distinguishable.
3) $0 \leq n(\epsilon) \leq 1$ for all single particle states with energy $\epsilon$ regardless of whether the particles are bosons or fermions.

A. 2 only    B. 3 only    **C. 1 and 2 only**    D. 2 and 3 only    E. All of the above.

*Class Discussion*

$D(\epsilon)$ - "Density of states" - The number of single particle states per unit energy with energy $\epsilon$.
$n(\epsilon)$ - "Distribution Function" – The average number of particles in a given single particle state with energy $\epsilon$.
$N(\epsilon) = D(\epsilon)n(\epsilon)$ is the average number of particles per unit energy with energy $\epsilon$.

Depending on the type of particles, the distribution function $n(\epsilon)$ can be one of the following:
- Maxwell-Boltzmann: $n(\epsilon) = \dfrac{1}{e^{\frac{\epsilon-\mu}{k_B T}}}$
- Fermi-Dirac: $n(\epsilon) = \dfrac{1}{e^{\frac{\epsilon-\mu}{k_B T}}+1}$
- Bose-Einstein: $n(\epsilon) = \dfrac{1}{e^{\frac{\epsilon-\mu}{k_B T}}-1}$

Review the differences between <u>fermions</u>, <u>bosons</u>, and <u>distinguishable particles</u> with respect to issues discussed in the two prior questions.

---

***(CQ14)*** Choose all of the following statements that are correct about the Maxwell-Boltzmann distribution (MBD), $n(\epsilon) = \dfrac{1}{e^{\frac{\epsilon-\mu}{k_B T}}}$.

1) The Maxwell-Boltzmann distribution can be used for classical distinguishable particles.
2) In the expression $n(\epsilon) = \dfrac{1}{e^{\frac{\epsilon-\mu}{k_B T}}}$, $\epsilon$ represents the energy of a single-particle state.
3) In the high-T limit, the Fermi-Dirac and Bose-Einstein distribution functions reduce to the MBD $n(\epsilon) = \dfrac{1}{e^{\frac{\epsilon-\mu}{k_B T}}}$.

A. 1 only    B. 2 only    C. 1 and 2 only    D. 1 and 3 only    **E.** All of the above.

---

***(CQ15)*** Choose all of the following statements that are correct about non-interacting fermions. Recall that the Fermi-Dirac distribution function is $n(\epsilon) = \dfrac{1}{e^{\frac{\epsilon-\mu}{k_B T}}+1}$.

1) At $T = 0$ K (absolute zero temperature), $n(\epsilon)=1$ if $\epsilon > \mu(T=0)$ and $n(\epsilon)=0$ if $\epsilon < \mu(T=0)$.
2) At $T = 0$ K (absolute zero temperature), the Fermi energy is equal to the chemical potential $\mu(T=0)$.
3) At a finite non-zero temperature, if $\epsilon = \mu(T)$, the average occupation number for a particular single particle state with energy $\epsilon$ is $n(\epsilon) = 1/2$.

A. 1 only    B. 1 and 2 only    C. 1 and 3 only    **D.** 2 and 3 only    E. All of the above.

***(CQ16)*** Choose all of the following statements that are correct about non-interacting bosons.

(1) The chemical potential for a bosonic system is always less than or equal to zero, $\mu(T) \leq 0$ (assuming $\epsilon \sim 0$ corresponds to the ground state).
(2) As the temperature decreases, the chemical potential $\mu(T)$ increases.
(3) Bose-Einstein condensation can occur at low temperatures when a macroscopic number of bosons occupies the lowest single-particle state or ground state.

A. 1 only    B. 1 and 2 only    C. 1 and 3 only    D. 2 and 3 only    **E.** All of the above.

*(CQ17)* Given the distribution function $n(\epsilon) = \frac{1}{e^{\frac{\epsilon-\mu(T)}{k_B T}} - 1}$ for massive bosons, choose all of the following statements that are true related to the chemical potential of the system:

1) $n(\epsilon)$ cannot be negative, so $\frac{\epsilon-\mu(T)}{k_B T} > 0$.
2) $\epsilon - \mu(T) > 0$ implies that $\epsilon > \mu(T)$ for all allowed single particle energies $\epsilon$.
3) Since the lowest single particle energy $\epsilon \sim 0$, $\epsilon > \mu(T)$ for all $\epsilon$ implies that $\mu(T)$ for a boson is always negative.

A. 1 only   B. 1 and 2 only   C. 1 and 3 only   D. 2 and 3 only   **E**. All of the above.

---

*Class Discussion*

*Discuss issues pertaining to the preceding questions such as these:*

*At high temperature, why do the Fermi-Dirac and Bose-Einstein distribution functions both reduce to the Maxwell-Boltzmann distribution function?*

*What do the graphical representations of the various distribution functions look like at different temperatures?*